\documentclass[a4paper,11pt]{article}
\usepackage{jinstpub}

\usepackage{graphicx}
\usepackage{subcaption}
\usepackage{amsmath}
\usepackage{hyperref}
\usepackage{xspace}
\usepackage{siunitx}
\usepackage{booktabs}
\usepackage{placeins}

\title{Sensor Quality Control and Annealing Studies of HGCAL Silicon Sensors}

\author[a,b]{Gizem Gul Dincer}
\author{on behalf of the CMS HGCAL Collaboration}

\affiliation[a]{CERN, Geneva, Switzerland}
\affiliation[b]{Karlsruhe Institute of Technology, Karlsruhe, Germany (DE)}

\emailAdd{gizem.gul.dincer@cern.ch}

\abstract{%
We summarise Sensor Quality Control (SQC) results of non-irradiated silicon sensors for the CMS HGCAL detector, as well as the first detailed annealing campaign with a wafer-scale 120\,\textmu m (Epitaxial) sensor exposed to \(2\times10^{15}\)\,\si{n_{eq}/cm^2}
at the Rhode Island Nuclear Science Center (RINSC).
For the non-irradiated sensors, we present an overview of the QC workflow developed
for HGCAL, including automated handling of vendor data, validation of electrical
measurements, and cross-checking of wafer-level characteristics.
The study investigates, for the first time, the isothermal annealing behaviour at
60\,\si{\celsius} after annealing periods ranging from 10 to 5000 minutes.
Hamburg-model parameters for effective doping concentration changes with annealing time, extracted from full-sensor data,
are presented.
The post-irradiation behaviour of sensors with hot regions in the pre-irradiation
leakage current measurements, as well as epitaxial sensors with stacking faults in individual
cells, is also investigated.%
}


\begin{document}
\maketitle

\section{Introduction}
\label{sec:intro}

\begin{figure}[htbp]
  \centering

  \begin{subfigure}[b]{0.37\textwidth}
    \centering
    \includegraphics[width=\textwidth]{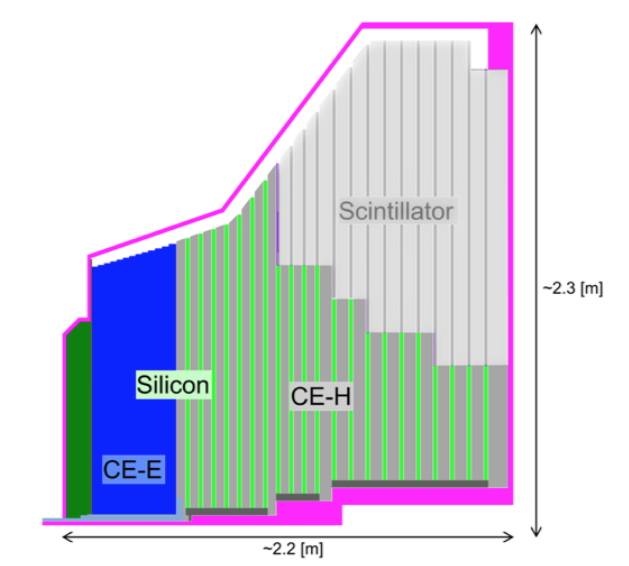}
    \caption{}
    \label{fig:hgcal_structure}
  \end{subfigure}
  \hspace{0.001\textwidth} 
  \begin{subfigure}[b]{0.49\textwidth}
    \centering
    \includegraphics[width=\textwidth]{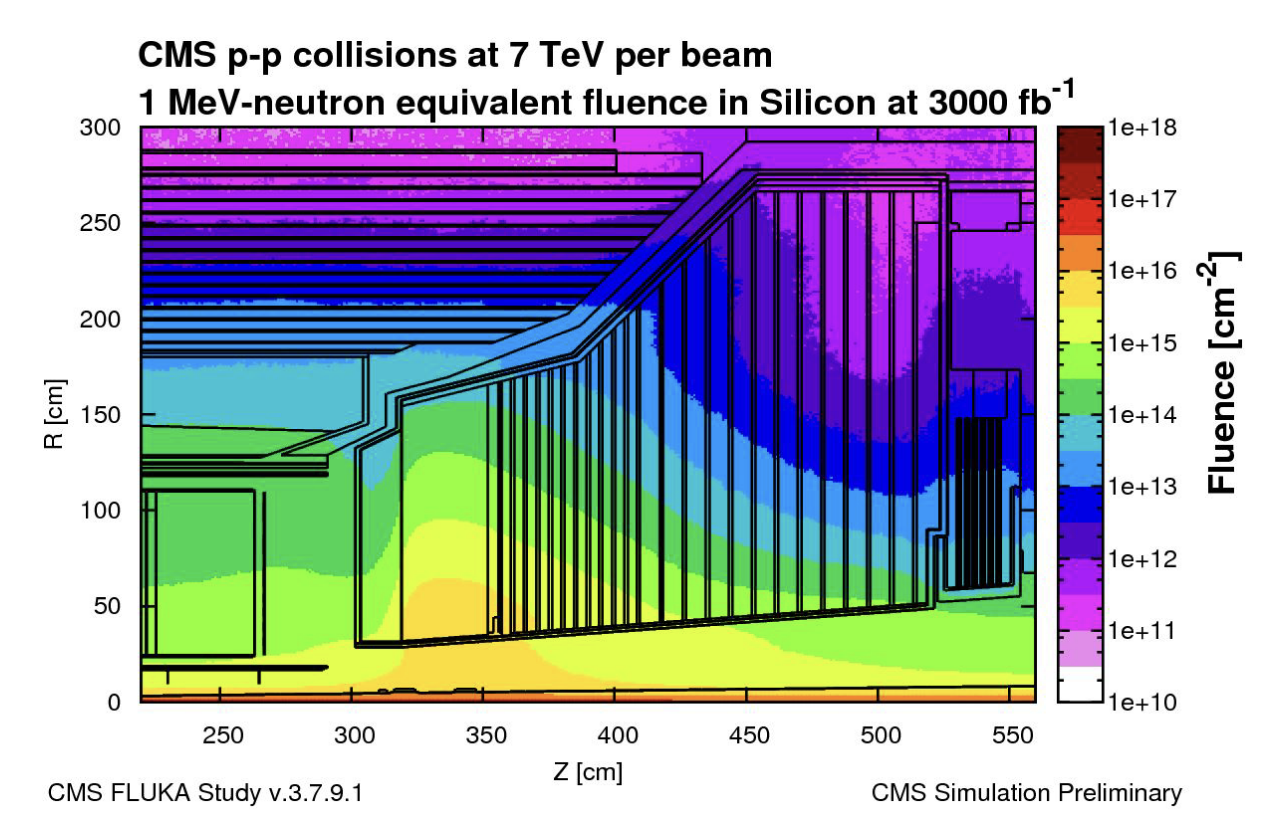}
    \caption{}
    \label{fig:fluence_map}
  \end{subfigure}

  \caption{(a) The CMS HGCAL structure~\cite{barney2022} and (b) expected fluence after ten years of operation~\cite{cms-tdr}.}
  \label{fig:radiation_env}
\end{figure}

The CMS detector at the CERN Large Hadron Collider is undergoing major upgrades for
the High-Luminosity LHC (HL-LHC), including the replacement of its calorimeters endcap (CE)
with the High Granularity Calorimeter (HGCAL). HGCAL is divided into two main calorimetric
regions: the Electromagnetic section (CE-E) and the Hadronic section (CE-H). It operates at \(-35\)\,\si{\celsius} to suppress radiation-induced leakage current in HGCAL's active materials.
The CE-E uses only silicon sensors as active material, whilst the CE-H region uses silicon sensors in
the regions exposed to highest radiation, and Scintillating Tiles with on-tile  Silicon Photomultiplier (SiPM) readout in the regions of lower radiation.
 Figure~\ref{fig:radiation_env} shows (\ref{fig:hgcal_structure}) the structure of HGCAL~\cite{barney2022} and (\ref{fig:fluence_map}) the expected neutron fluence over ten years of operation~\cite{cms-tdr}. The system is designed to endure particle fluences up to \(1.5\times10^{16}\)\,\si{n_{eq}/cm^2}
and radiation doses reaching 1.5\,MGy over ten years of operation. The HGCAL uses 8-inch p-type silicon sensors, selected for their durability under
the extreme radiation expected during HL-LHC operation. The sensors are divided into smaller hexagonal cells of around 1.1\,cm$^2$ or
0.5\,cm$^2$ and assembled into modules. More detailed information on the HGCAL layout can be found in~\cite{cms-tdr}.

\section{HGCAL 8-inch Silicon Sensor Tests}
\label{sec:sensors}
\begin{figure}[htbp]
  \centering
  \includegraphics[width=0.40\textwidth]{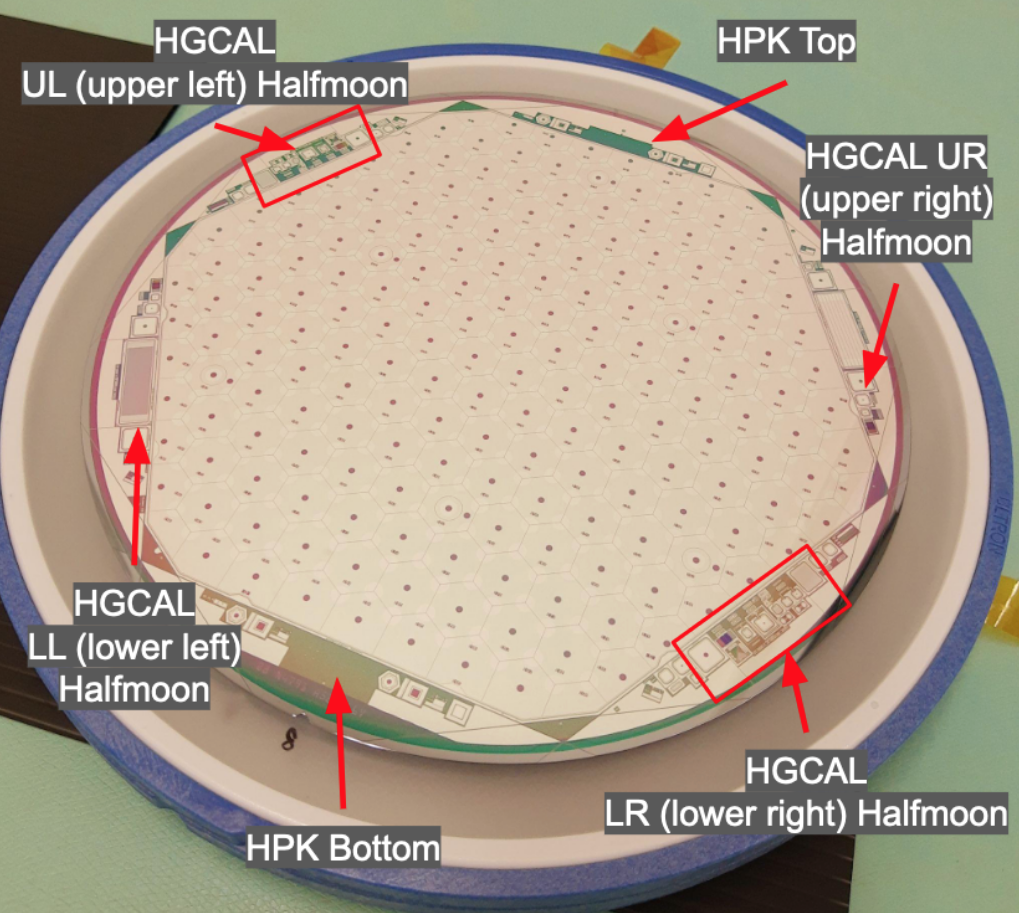}
  \caption{Full-wafer silicon sensor in a dicing frame, showing the main hexagonal sensor at the centre surrounded by half-moon test structures, with red boxes indicating those used for PQC measurements~\cite{kaluzinska2022}.}
  \label{fig:halfmoon}
\end{figure}

\vspace{-6pt}
The silicon sensors are fabricated on 8-inch p-type silicon wafers by Hamamatsu Photonics
K.K. (HPK) in Japan~\cite{hamamatsu}. Figure~\ref{fig:halfmoon}  shows a typical wafer, including the main, hexagonal sensor area and associated test structures.
Optimised for the radiation field, sensors are produced with three active thicknesses:
300\,\textmu m, 200\,\textmu m (both float-zone process, FZ), and 120\,\textmu m
(epitaxial process, Epi). Thinner sensors are used in the higher-radiation regions of the calorimeter, as
shown in Figure~\ref{fig:radiation_env} (orange region, up to \(10^{16}\)\,\si{n_{eq}/cm^2}).
Depending on sensor thickness, there are two layout types: low-density (LD) and
high-density (HD), which define the number of cells~\cite{cms-tdr}.
The main hexagonal sensor is diced from the circular wafer, as shown in
Figure~\ref{fig:halfmoon}.
The remaining pieces, called ``halfmoons'' contain several test structures, such as single diodes, which share
the same production process as a cell in the full sensor~\cite{hinger2021}.

The manufacturer completed the production of nearly 25\,000
8-inch p-type silicon wafers in the time starting from February 2023 to May 2025. This large-scale production included per-cell leakage current (IV) and capacitance
(CV) measurements, as well as optical inspections for mechanical defects performed both at the vendor and
during subsequent acceptance stages at multiple CMS institutes~\cite{hinger2021}.

\section{Quality Assurance for HGCAL Silicon Sensors}
\label{sec:qa}
\vspace{-6pt}

There are four steps of batch-based qualification for the silicon sensors~\cite{hinger2021}.
All sensors are tested by HPK, a step called Vendor Quality Control (VQC). HPK then transmits the accepted sensors to the
distribution centre at CERN, along with all corresponding test structures.
All sensors and corresponding test structures are then transferred from the
distribution centre at CERN to the relevant CMS test centres.
These institutes carry out Sensor Quality Control (SQC) on the sensors and Process
Quality Control (PQC) on the test structures. About 8\% of the sensors are tested for SQC
and about 4\% for PQC, corresponding to one or two sensor per batch of up to 25 sensors.
An additional 0.1\% of the sensors undergo irradiation tests (IT). Electrical and optical tests provide information on the technical characteristics of the sensor.
Full-wafer IV\,+\,CV measurements are performed at all SQC institutes equipped with
ARRAY systems~\cite{pitters2019}. Sensors are rejected based on the leakage current specifications as defined in~\cite{dincer2024}.

\begin{figure}[htbp]
  \centering

  \begin{subfigure}[b]{0.48\textwidth}
    \centering
    \includegraphics[width=\textwidth]{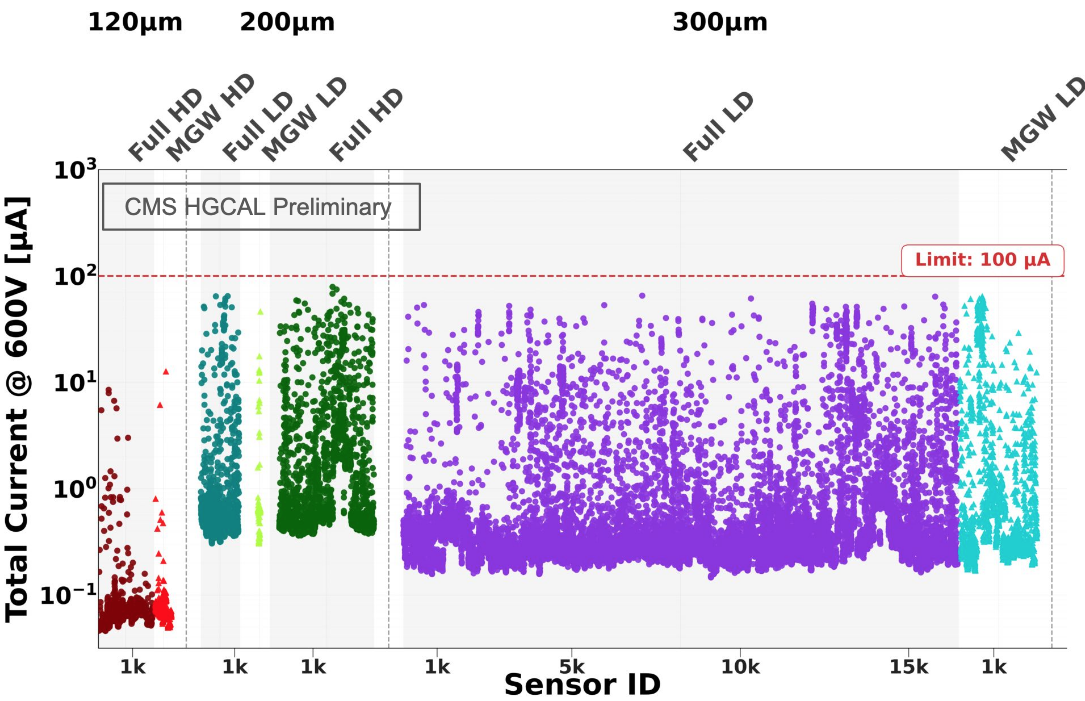}
    \caption{}
    \label{fig:current_600V}
  \end{subfigure}
  \hspace{0.001\textwidth} 
  \begin{subfigure}[b]{0.47\textwidth}
    \centering
    \includegraphics[width=\textwidth]{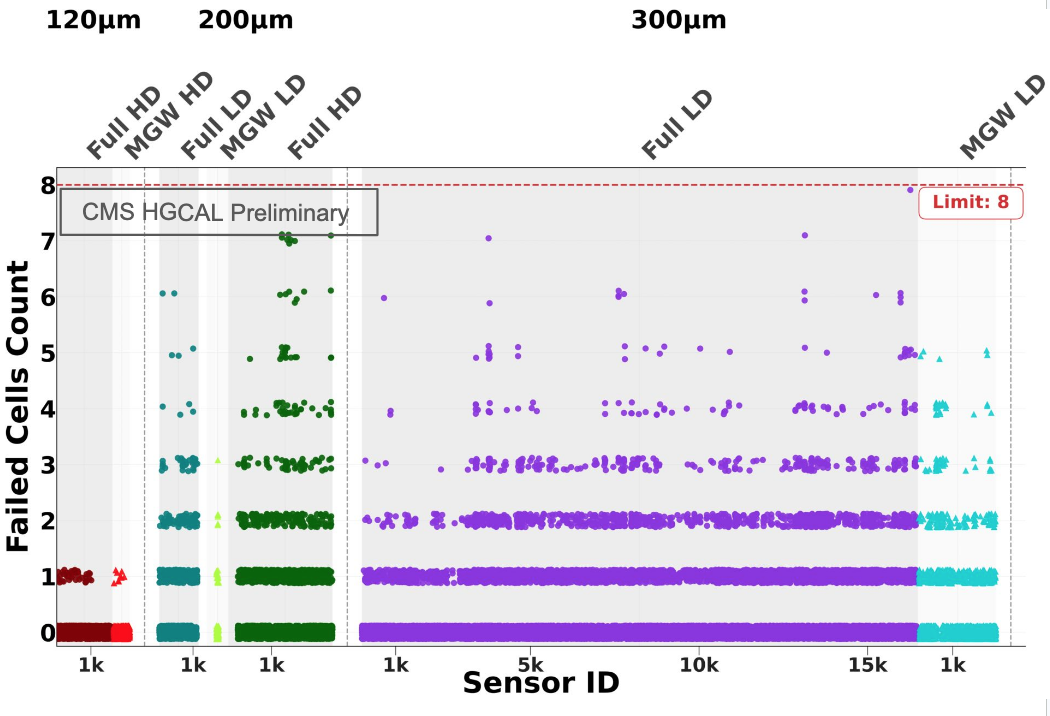}
                        \caption{}
                        \label{fig:failed_cells}
  \end{subfigure}
\vspace{-6pt}
  \caption{Total current at 600\,V\,(a) and number of failed cells\,(b) distribution
    for all sensor types and thicknesses (Full LD/HD and MGW LD/HD).}
  \label{fig:qc_summary}
\end{figure}

Figure~\ref{fig:qc_summary} summarises the total current at 600\,V and the number
of failed cells for each sensor thickness and type (full and partial, which are obtained
from Multi-Geometry Wafers, MGW) using the HPK-supplied data. We observe that all delivered sensors have less than 100\,\textmu A total current, and that
all but one sensor have fewer than 8 bad cells. During quality inspections, several types of optical and electrical defects have been observed.

\vspace{-6pt}

The most common optical defects include stacking faults and scratches on the sensor surface.
We found that bad cells in epitaxial sensors are often linked to stacking faults~\cite{airaksinen2005} formed during the manufacturing process, which can be easily recognised during optical inspection. For the SQC step, electrical tests were performed on 1950 sensors ($\sim$8\%)  --- 60 with visible damage and 25 (1.28\%) with non-compliant IV results --- whilst optical inspections were carried out more frequently --- reaching 50\% of the total.

\section{Irradiation Studies}
\label{sec:irradiation}

\subsection{Neutron Irradiation at RINSC and annealing setup}
\label{sec:rinsc}

The sensors are irradiated with neutrons at the Rhode Island Nuclear Science Center
(RINSC), the only irradiation facility capable of accommodating 8-inch wafers~\cite{rinsc1}. The sensors are exposed to high temperatures during irradiation in the reactor;
aluminium cylinders filled with dry ice limit heating of the sensors.
Resistance Temperature Detectors (RTDs) are employed to record the temperature during
irradiation. These RTD measurements are taken into account in the annealing steps performed at CERN as well as any annealing time experienced in the reactor.

\begin{figure}[htbp]
  \centering

  \begin{subfigure}[b]{0.22\textwidth}
    \centering
    \includegraphics[width=\textwidth]{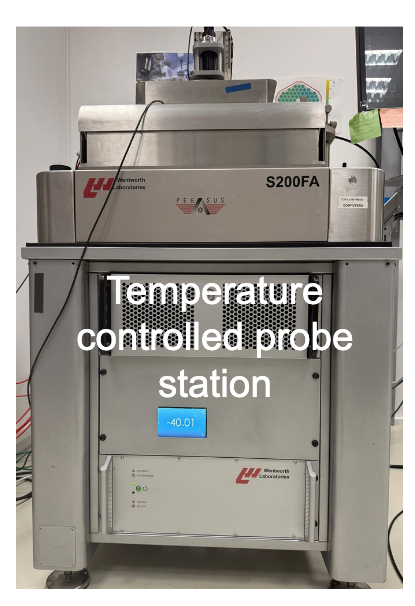}
    \caption{}
    \label{fig:probe_station_closed}
  \end{subfigure}
  \hspace{0.02\textwidth} 
  \begin{subfigure}[b]{0.37\textwidth}
    \centering
    \includegraphics[width=\textwidth]{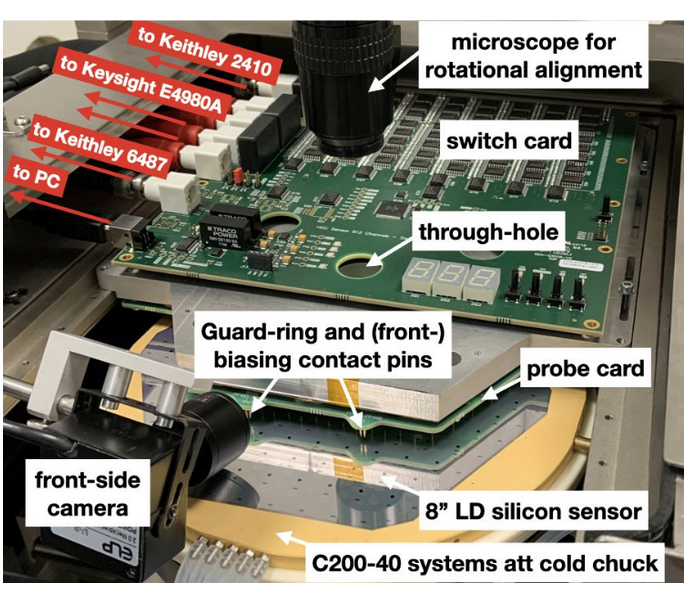}
    \caption{}
    \label{fig:array_system}
  \end{subfigure}
\vspace{-6pt}
  \caption{(a) The temperature-controlled probe station when it is closed.
  (b) The ARRAY system and the silicon sensor placed on the temperature controlled chuck inside the probe station~\cite{rinsc1}.}
  \label{fig:alps}
\end{figure}

In this study, the electrical characterisation of a 120\,\textmu m full-wafer
production  sensor (ID\,=\,300272), irradiated to a fluence of
\(2\times10^{15}\)\,\si{n_{eq}/cm^2}, is investigated. The in-reactor annealing time was calculated to be approximately 10 minutes, and
this duration was added to the subsequent annealing times at CERN. After irradiation we performed 15 annealing steps at 60\,\si{\celsius} for a total of approximately
5000 minutes using a temperature controlled chuck of a probe station (Figure~\ref{fig:alps}). IV/CV
measurements at $-35$\,\si{\celsius} of all cells were performed after each annealing step.

\subsection{Saturation Voltages from CV Measurements}
\label{sec:cv}

\begin{figure}[htbp]
  \centering

  \begin{subfigure}[b]{0.40\textwidth}
    \centering
    \includegraphics[width=\textwidth]{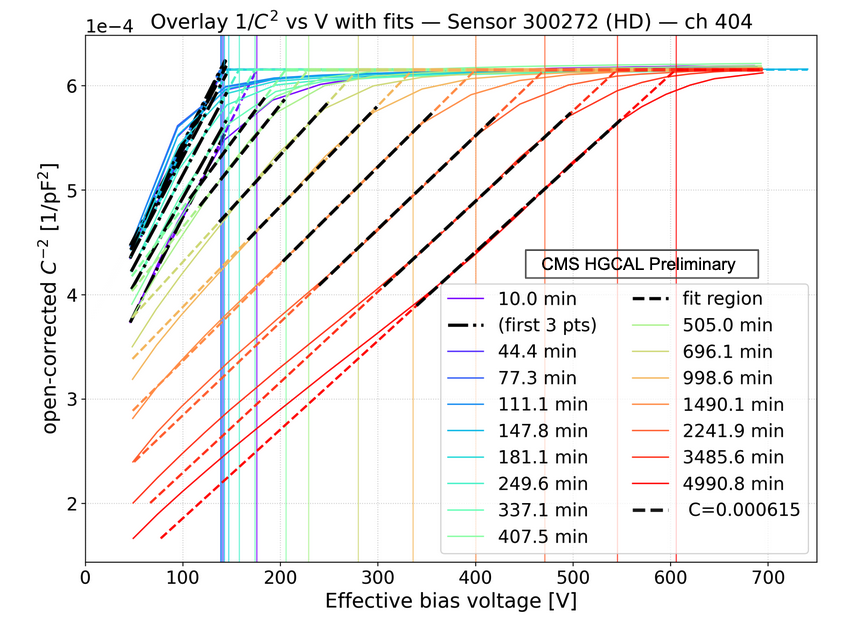}
    \caption{}
    \label{fig:cv_channel404}
  \end{subfigure}
  \hspace{0.001\textwidth} 
  \begin{subfigure}[b]{0.45\textwidth}
    \centering
    \includegraphics[width=\textwidth]{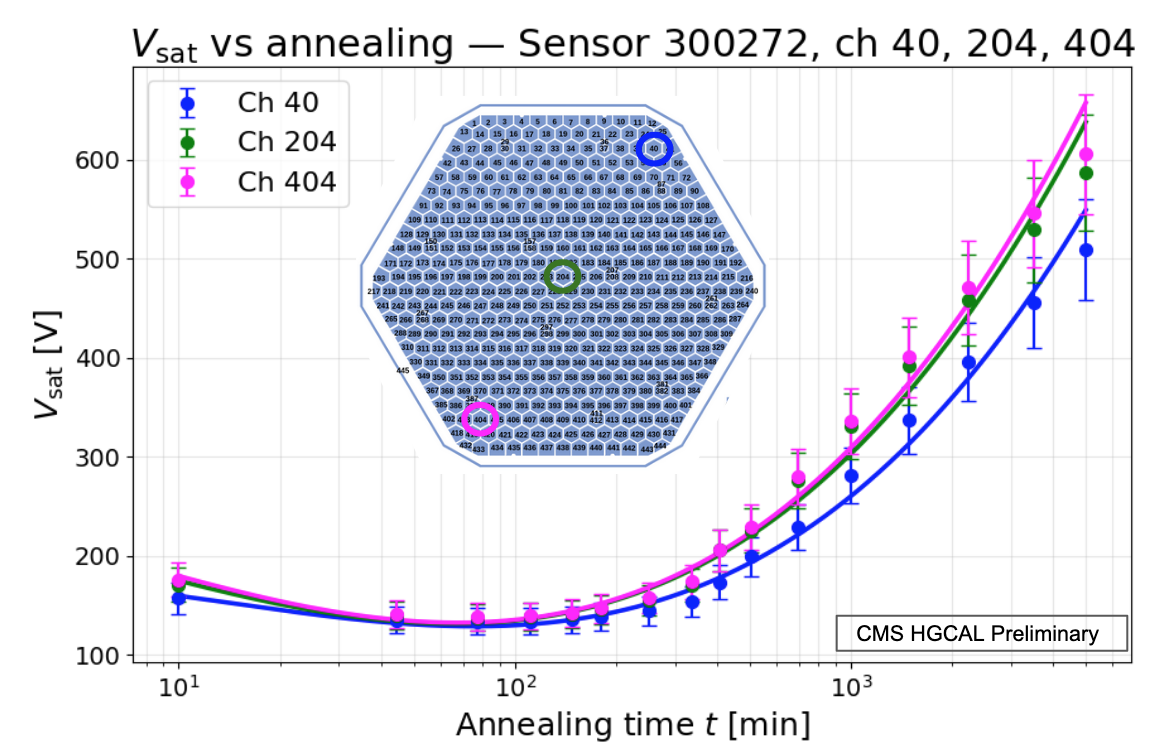}
    \caption{}
    \label{fig:sat_voltage}
  \end{subfigure}
\vspace{-6pt}
  \caption{(a) The inverse of the capacitance squared versus the bias voltage for
  different annealing times, for an example channel 404.
  (b) Variation in saturation voltages of sample channels 40, 204, and 404
  taken from three different regions of the sensor as a function of annealing time.}
  \label{fig:combined}
\end{figure}

After each annealing step the capacitance was measured as a function of the bias voltage  up to 700\,V.  Figure~\ref{fig:cv_channel404} shows the CV measurement for an example channel 404
 (one of the 444 channels), comparing the inverse of the capacitance squared with the
bias voltage for different annealing times.
At low annealing times, $1/C^2$ rises rapidly; at longer annealing times, it rises
more slowly. During long annealing times there are very few or no measurement points in the plateau region. The saturation voltage is estimated by the intersection point of two fits, one for rising part of the curve and one for the plateau~\cite{diode}.
For data sets where the plateau is not reached, the average of
the first three annealing steps is used.
An uncertainty of 10\% on the estimated saturation voltage has been assumed.

For the fit of the rising part of the $1/C^2$ vs. $V$ curve, at low annealing times first three voltage points are taken into account while at high annealing times, $1/C^2$ values between 50\%--90\% of plateau value  were
included in the fit. Figure~\ref{fig:sat_voltage} shows the variation in saturation voltages of example
channels from three different regions of the sensor as a function of annealing time. In future campaigns, finer voltage steps will be used to yield a better sampling of the dependence. Different saturation voltage trends were observed between the different sensor regions, which are further discussed in sections~\ref{sec:hexaplot} and~\ref{sec:neff}.

\subsection{Annealing Behaviour: Per-Cell Leakage Current}
\label{sec:leakage}

\begin{figure}[htbp]
  \centering

  \begin{subfigure}[b]{0.44\textwidth}
    \centering
    \includegraphics[width=\textwidth]{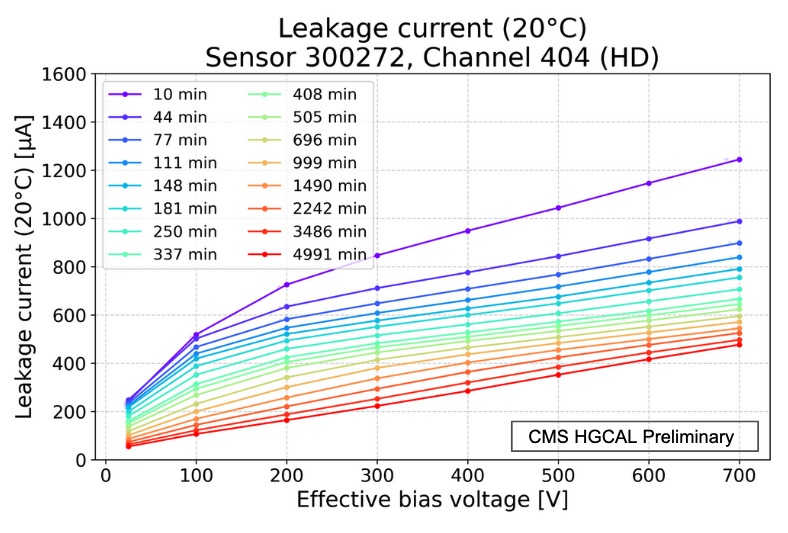}
    \caption{}
    \label{fig:leakage_ch404}
  \end{subfigure}
  \hspace{0.001\textwidth} 
  \begin{subfigure}[b]{0.45\textwidth}
    \centering
    \includegraphics[width=\textwidth]{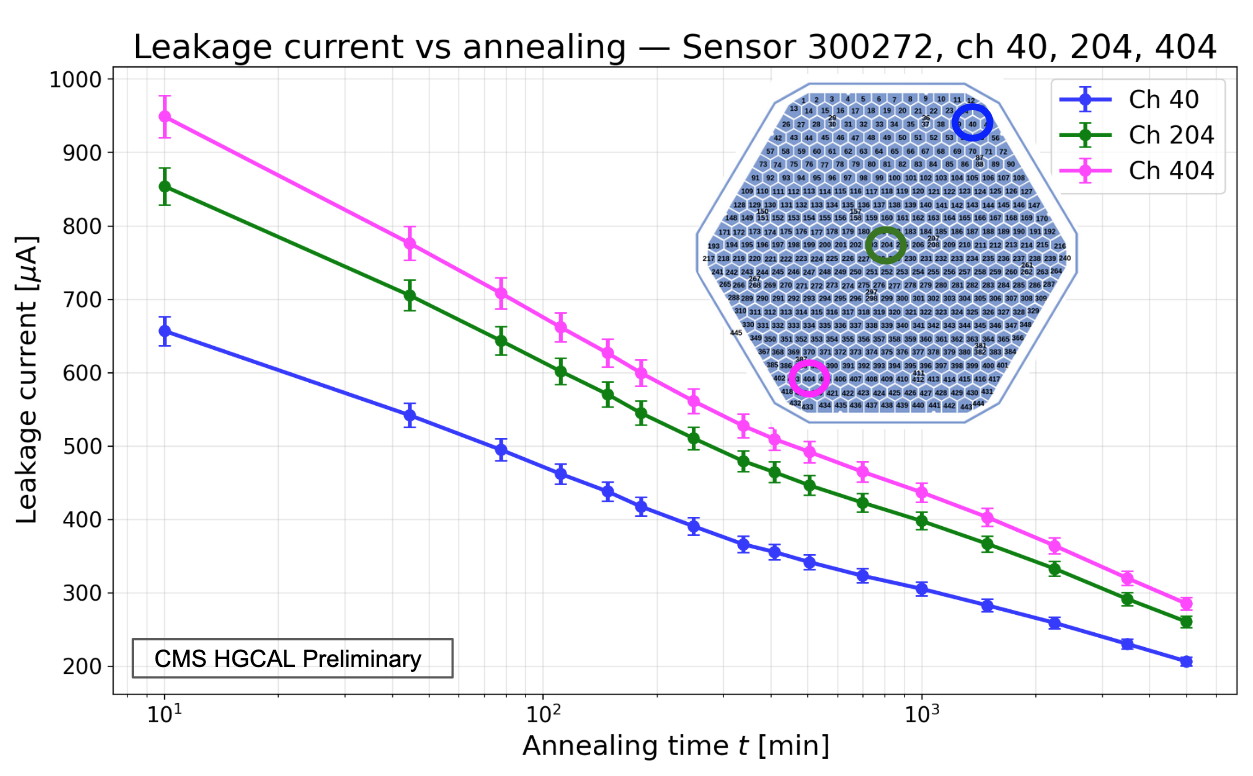}
    \caption{}
    \label{fig:leakage_scaled}
  \end{subfigure}
\vspace{-6pt}
  \caption{(a) The leakage current measured for example channel 404 after each annealing
  step as a function of bias voltage.
  (b) Measured leakage current at 400\,V, scaled to 20\,\si{\celsius} using
  Arrhenius scaling, versus annealing time for three channels from three different
  sensor regions.}
  \label{fig:leakage_combined}
\end{figure}

The leakage current was measured for all channels after each annealing step as a
function of the bias voltage up to 700\,V.
As shown in Figure~\ref{fig:leakage_ch404}, for example channel 404, the leakage
current decreases with annealing time for all voltage steps. 

In Figure~\ref{fig:leakage_scaled}, the measured leakage current $I$ at 400\,V is scaled to $20$\,\si{\celsius} using the Arrhenius scaling and used to calculate the current related annealing behaviour for three channels from three different regions of the sensor. The leakage current decreases as expected with annealing time for all channels. However, the cells from the upper-right region of the sensor have less leakage current than
the middle and bottom-left region cells, which is further discussed in section~\ref{sec:hexaplot}.

\subsection{Per-Cell Leakage Current and Saturation Voltage}
\label{sec:hexaplot}

\begin{figure}[htbp]
  \centering

  \begin{subfigure}[b]{0.40\textwidth}
    \centering
    \includegraphics[width=\textwidth]{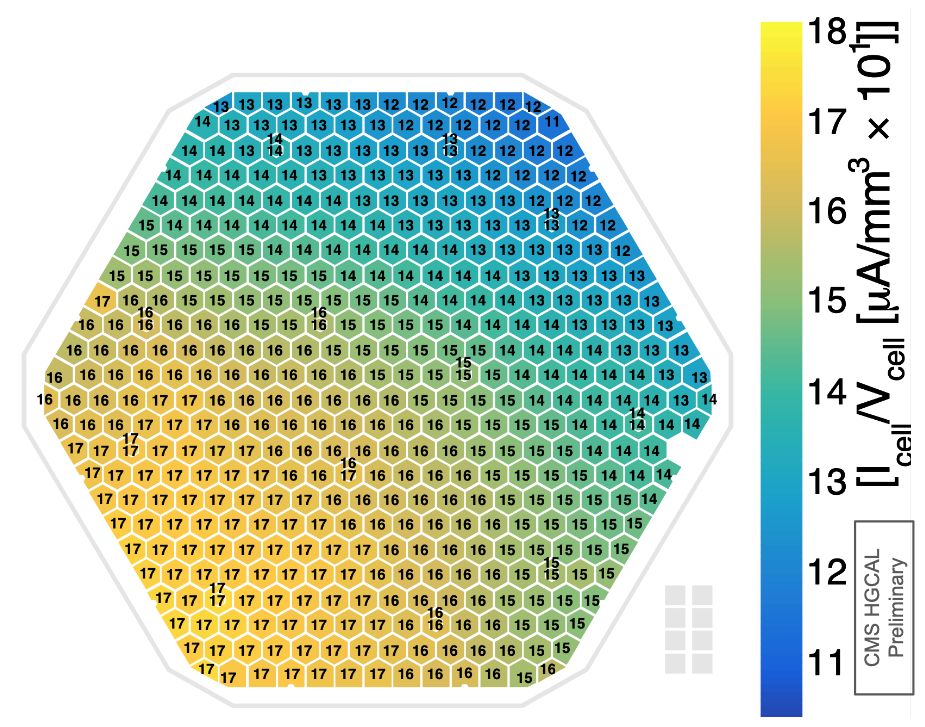}
    \caption{}
    \label{fig:leakage_600V}
  \end{subfigure}
  \hspace{0.02\textwidth} 
  \begin{subfigure}[b]{0.40\textwidth}
    \centering
    \includegraphics[width=\textwidth]{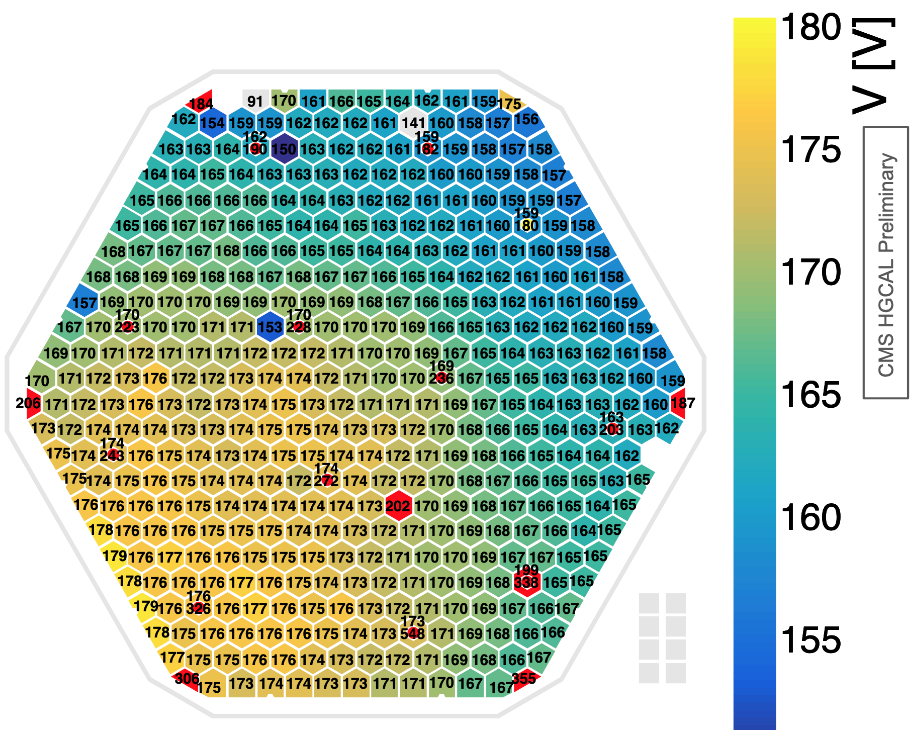}
    \caption{}
    \label{fig:depletion_voltage}
  \end{subfigure}
\vspace{-6pt}
  \caption{Volume-normalised per-cell leakage current at 600\,V\,(a) and depletion voltage (b) after 10\,min reactor annealing time.}
  \label{fig:hexaplot}
\end{figure}

Figure~\ref{fig:hexaplot} shows 2D distributions of the volume-normalised per-cell leakage current at 600\,V (\ref{fig:leakage_600V})
and the depletion voltage fit results (\ref{fig:depletion_voltage}) after 10\,min reactor annealing time.


The leakage current values are 30\% lower and the saturation voltages are approximately
10\% lower in the upper-right region than in the bottom left region.
This different behaviour of leakage current and saturation voltages in different
regions of the full sensor was observed previously and was interpreted as an asymmetric fluence profile of the irradiation at RINSC~\cite{rinsc1,rinsc2}. This is further evaluated in section~\ref{sec:neff}.

\vspace{-14pt}

\vspace{-6pt}

\subsection{Hamburg Model Fit: Effective Doping Concentration}
\label{sec:neff}

\begin{figure}[htbp]
  \centering

  \begin{subfigure}[b]{0.40\textwidth}
    \centering
    \includegraphics[width=\textwidth]{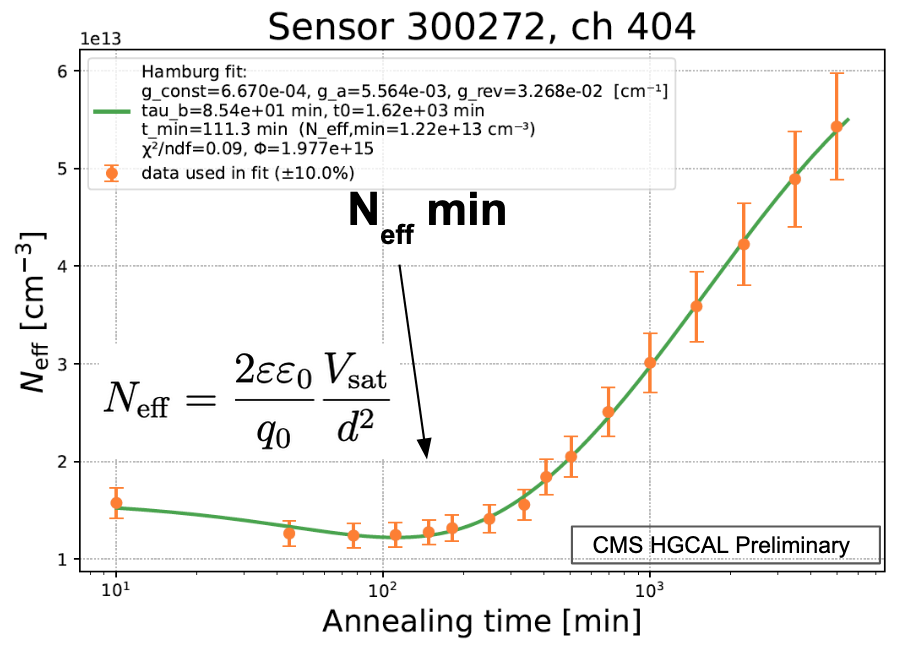}
    \caption{}
    \label{fig:neff_channel404}
  \end{subfigure}
  \hspace{0.001\textwidth} 
  \begin{subfigure}[b]{0.40\textwidth}
    \centering
    \includegraphics[width=\textwidth]{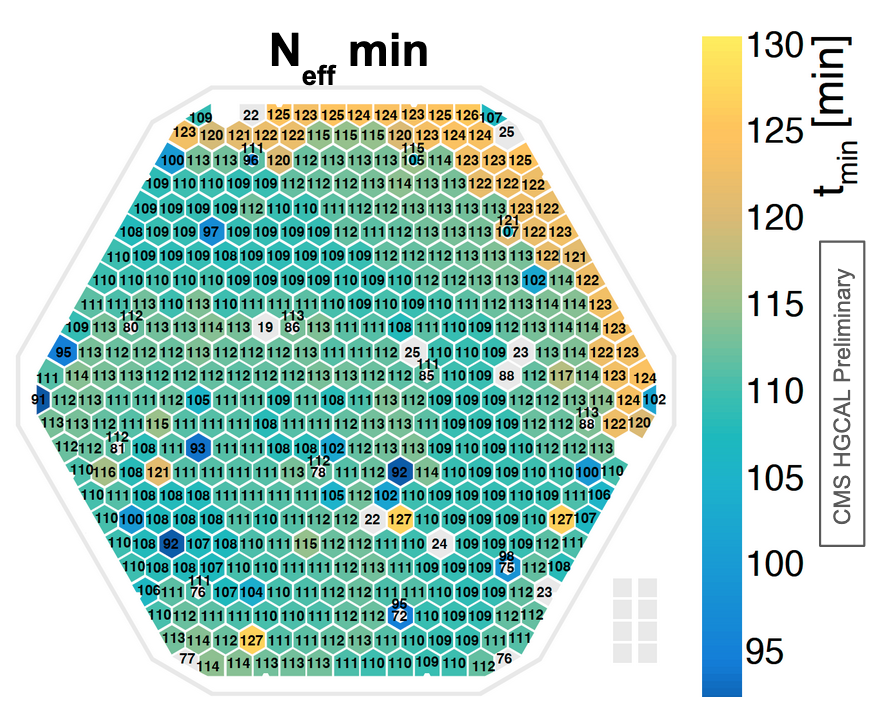}
    \caption{}
    \label{fig:neff_hexaplot}
  \end{subfigure}
\vspace{-8pt}
  \caption{Effective doping concentration versus annealing time for example
    channel 404\,(a) and minimum $N_{\mathrm{eff}}$ per channel hexaplot\,(b).}
  \label{fig:neff}
  \vspace{-6pt}
\end{figure}

The effective doping concentration $N_{\mathrm{eff}}$ is related to the saturation
voltage $V_{\mathrm{sat}}$ by:
\begin{equation}
  N_{\mathrm{eff}} = \frac{2\,\varepsilon_0\,\varepsilon_r\,V_{\mathrm{sat}}}{q\,d^2},
  \label{eq:neff}
\end{equation}
where $d$ is the sensor thickness, $q$ the elementary charge, $\varepsilon_0$ the vacuum
permittivity, and $\varepsilon_r$ the relative permittivity of silicon.
Using equation~\eqref{eq:neff} we convert saturation voltage values for each channel
into $N_{\mathrm{eff}}$ (Figure~\ref{fig:neff_channel404}) and fit the minimum value of
$N_{\mathrm{eff}}$ for each channel, shown in the hexaplot in Figure~\ref{fig:neff_hexaplot} as extracted from the Hamburg model fit using equation~\eqref{eq:hamburg} introduced in section~\ref{sec:hamburg}.

As seen in Figure~\ref{fig:neff_hexaplot}, the minimum values of $N_{\mathrm{eff}}$ occur
approximately 10\,min later in the upper-right cells.
This 10\,min difference in annealing time would result in only $\sim$6\% leakage
current reduction, whereas the observed reduction is $\sim$30\%
(Figure~\ref{fig:leakage_600V}).
This suggests that the majority of the leakage current trend indeed originates from the
fluence profile as assumed in~\cite{rinsc1,rinsc2}.

\FloatBarrier

Based on this assumption, the fluence per cell is calculated using:
\vspace{-5pt}
\begin{equation}
  \Phi_{\mathrm{cell}} = \frac{(I/V)_{\mathrm{cell}}}{\max_{\mathrm{cells}}(I/V)} \cdot \Phi_{\mathrm{target}}
  \label{eq:alpha}
\end{equation}

\vspace{-2pt}
which is used in the following (Figure~\ref{fig:neff_third}):
\vspace{-12pt}
\begin{figure}[htbp]
  \centering
  \includegraphics[width=0.40\textwidth]{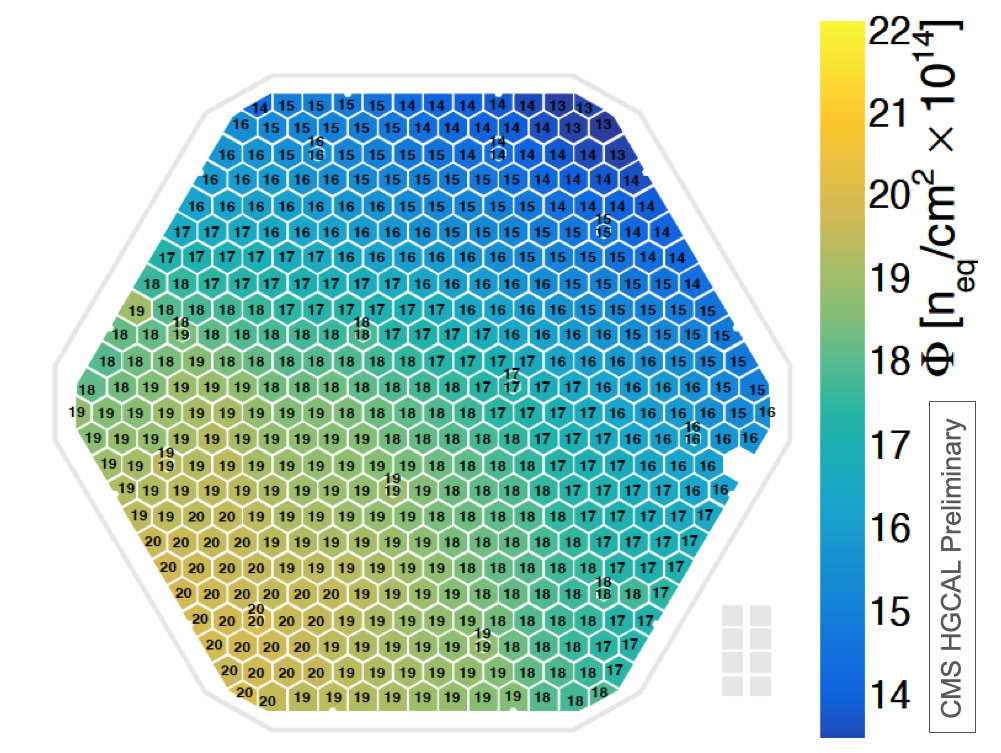}
  \vspace{-7pt}
  \caption{Fluence per cell is defined at first annealing step, only in-reactor annealing and used for all annealing times. The fluence range between 1.3 (top right) and 2.0 (bottom left)\(\times10^{15}\)\,\si{n_{eq}/cm^2}.}
  \label{fig:neff_third}
\end{figure}
\vspace{-16pt}
\subsection{Annealing Parameters from the Hamburg Model}
\label{sec:hamburg}
The Hamburg model for the effective doping concentration is described in Eq.~\eqref{eq:hamburg}~\cite{Moll:1999kv}. 

\vspace{-10pt}
\begin{equation}
  N_{\mathrm{eff}}(\Phi, t) = N_A(\Phi,t) + N_C (\Phi) + N_Y(\Phi,t) + N_{\mathrm{eff,0}}
  \label{eq:hamburg}
\end{equation}
\vspace{-12pt}
where
{\setlength{\jot}{3pt}
\begin{align}
  N_C &= g_c\,\Phi \quad \text{(stable damage)},\\
  N_A(t) &= g_a\,\Phi\,\exp(-t/\tau_a) \quad \text{(beneficial annealing)},\\
  N_Y(t) &= g_Y\,\Phi\,\left(1 - \frac{1}{1 + t/\tau_Y}\right)
           \quad \text{(reverse annealing)}.
\end{align}
}

\begin{figure}[htbp]
  \centering

  \begin{subfigure}[b]{0.37\textwidth}
    \centering
    \includegraphics[width=\textwidth]{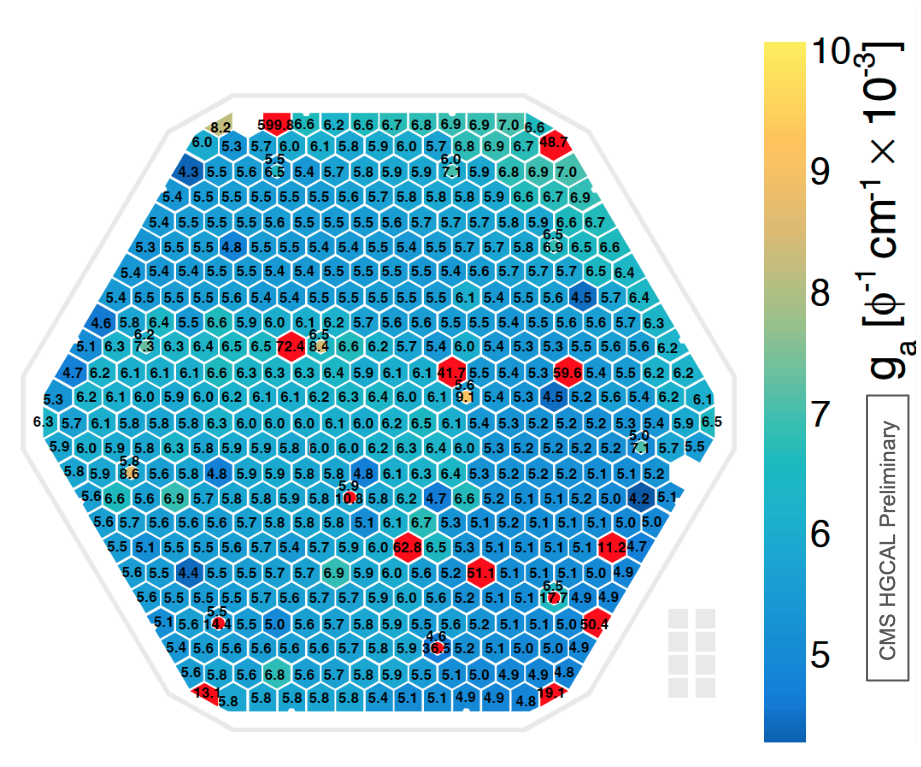}
    \caption{}
    \label{fig:ga_constant}
  \end{subfigure}
  \hspace{0.001\textwidth} 
  \begin{subfigure}[b]{0.37\textwidth}
    \centering
    \includegraphics[width=\textwidth]{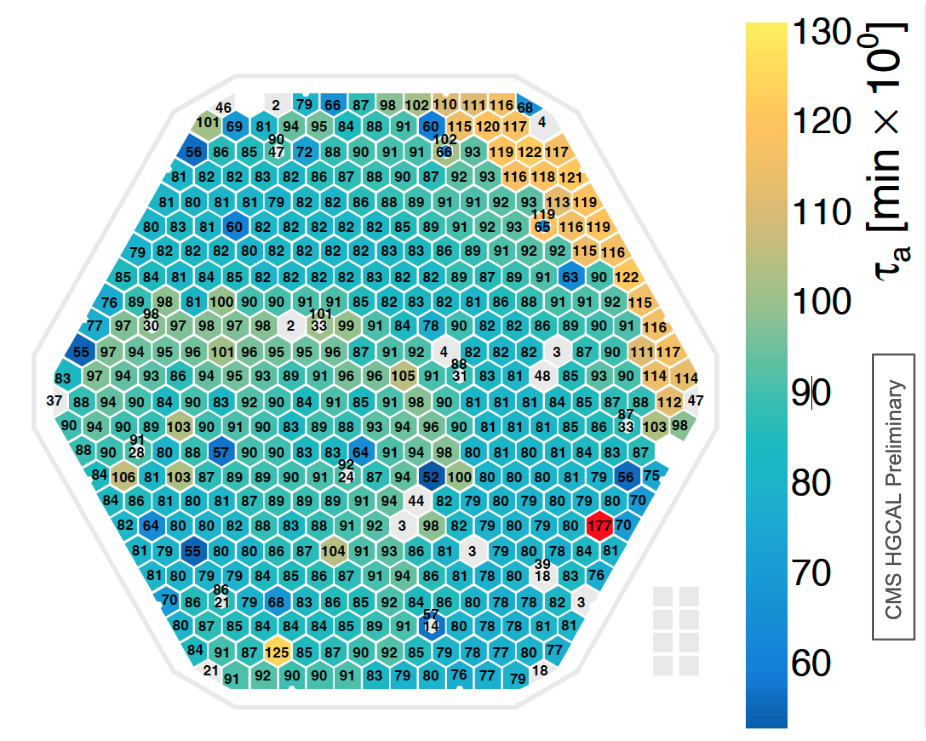}
    \caption{}
    \label{fig:tau_constant}
  \end{subfigure}
\vspace{-9pt}
  \caption{(a) The beneficial annealing constant $g_a$ for each channel and (b) its time
    constant $\tau_a$.}
  \label{fig:ga}
\end{figure}

\begin{figure}[htbp]
  \centering

  \begin{subfigure}[b]{0.37\textwidth}
    \centering
    \includegraphics[width=\textwidth]{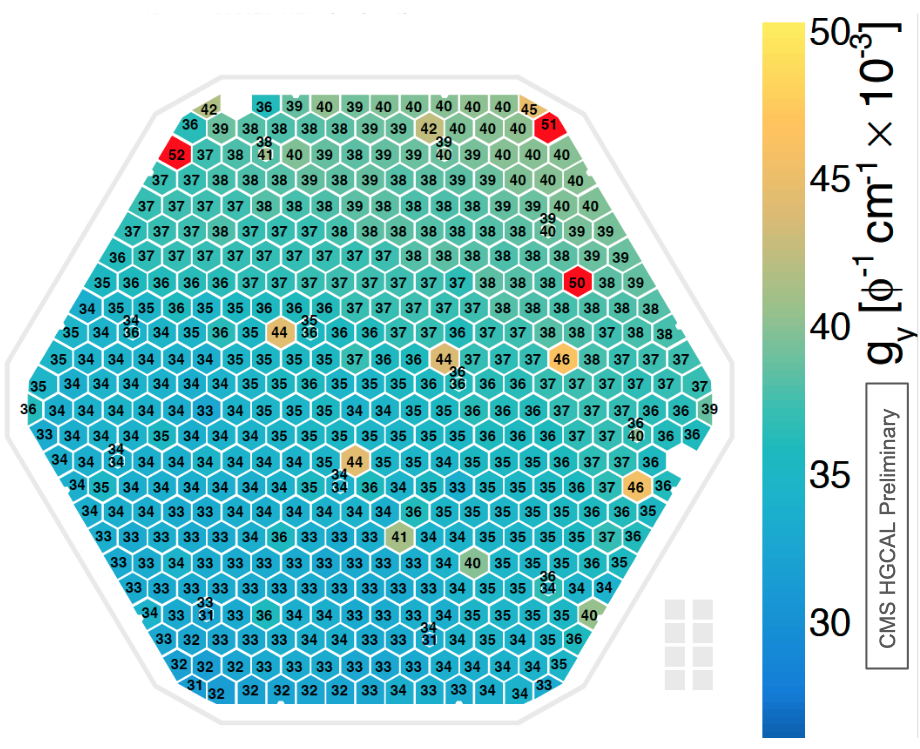}
    \caption{}
    \label{fig:gy_constant}
  \end{subfigure}
  \hspace{0.001\textwidth} 
  \begin{subfigure}[b]{0.37\textwidth}
    \centering
    \includegraphics[width=\textwidth]{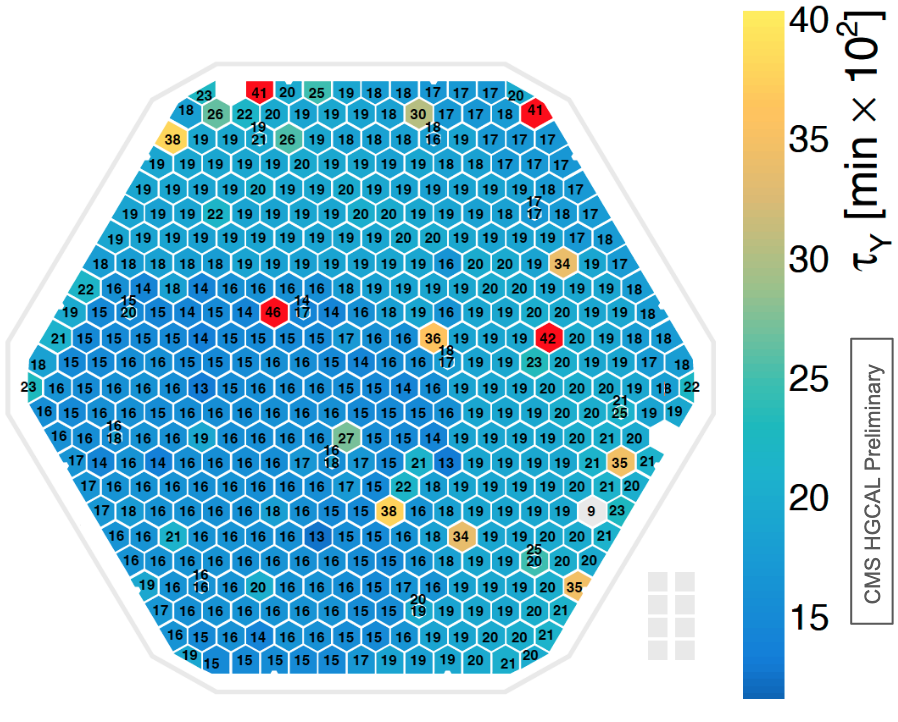}
    \caption{}
    \label{fig:tauY_constant}
  \end{subfigure}
\vspace{-9pt}
  \caption{(a) The reverse annealing constant $g_Y$ for each channel and (b) its time constant
    $\tau_Y$.}
  \label{fig:gy}
\end{figure}

\begin{figure}[htbp]
  \centering

  \begin{subfigure}[b]{0.37\textwidth}
    \centering
    \includegraphics[width=\textwidth]{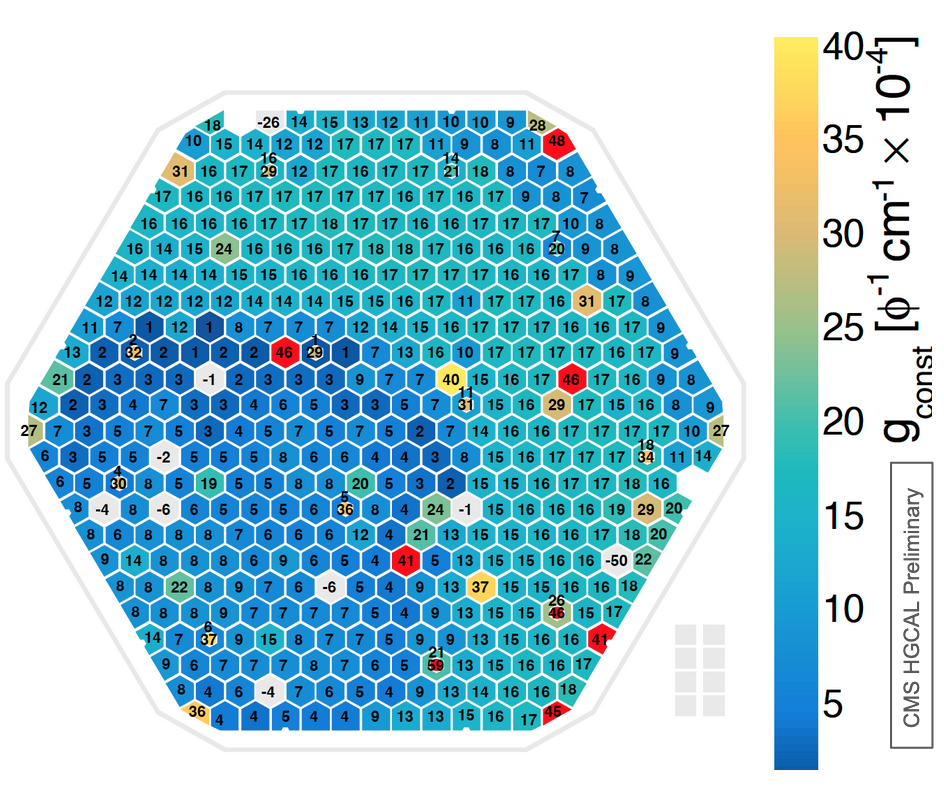}
    \caption{}
    \label{fig:gconst_channels}
  \end{subfigure}
  \hspace{0.001\textwidth} 
  \begin{subfigure}[b]{0.39\textwidth}
    \centering
    \includegraphics[width=\textwidth]{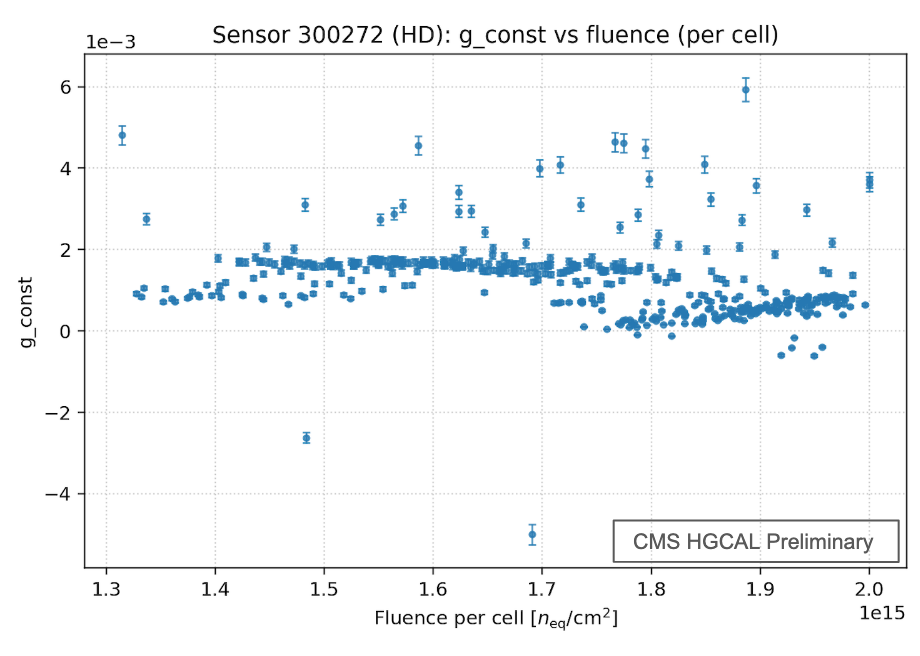}
    \caption{}
    \label{fig:gconst_distribution}
  \end{subfigure}
\vspace{-9pt}
  \caption{(a) The stable annealing constant $g_{\mathrm{const}}$ for each channel
    and (b) its distribution per fluence per cell.}
  \label{fig:gc}
\end{figure}

Fitting the $N_{\mathrm{eff}}$ values to the Hamburg model yields the annealing
constants. For the non-irradiated 120~$\mu$m sensor, an effective doping concentration of 
$N_{\mathrm{eff},0} = 3.75 \times 10^{12}\,\mathrm{cm}^{-3}$ was used~\cite{Kieseler2023}.
Figure~\ref{fig:ga} shows the beneficial annealing constant $g_a$ (\ref{fig:ga_constant}) and its time
constant $\tau_a$ (\ref{fig:tau_constant}). The beneficial annealing constant $g_a$ values decrease $\sim$20\%  with fluence (Figure~\ref{fig:ga_constant}) and
are mostly in agreement with results from single-diode irradiations~\cite{diode}. Figure~\ref{fig:gy} shows the reverse annealing constant $g_Y$ (\ref{fig:gy_constant}) and its time
constant $\tau_Y$ (\ref{fig:tauY_constant}). The reverse annealing constant $g_Y$ values also decrease  $\sim$15\%  with fluence (Figure~\ref{fig:gy_constant}),
in agreement with single-diode irradiation results~\cite{diode}. No dependence on the fluence is found for the stable damage parameter $g_c$ (Figure~\ref{fig:gc}).
Some of the $\sim$7000 CV fits (444 channels times 16 steps) did not converge correctly and give outliers.
\vspace{-11pt}

\vspace{-17pt}

\vspace{-20pt}

\vspace{-9pt}

The results represent the first detailed fluence dependence study of annealing time constants for 8-inch p-type silicon sensors.

\vspace{-3pt}
\section{Stacking Faults after Irradiation}
\label{sec:stacking}
\vspace{-6pt}

\begin{figure}[htbp]
  \centering

  \begin{subfigure}[b]{0.34\textwidth}
    \centering
    \includegraphics[width=\textwidth]{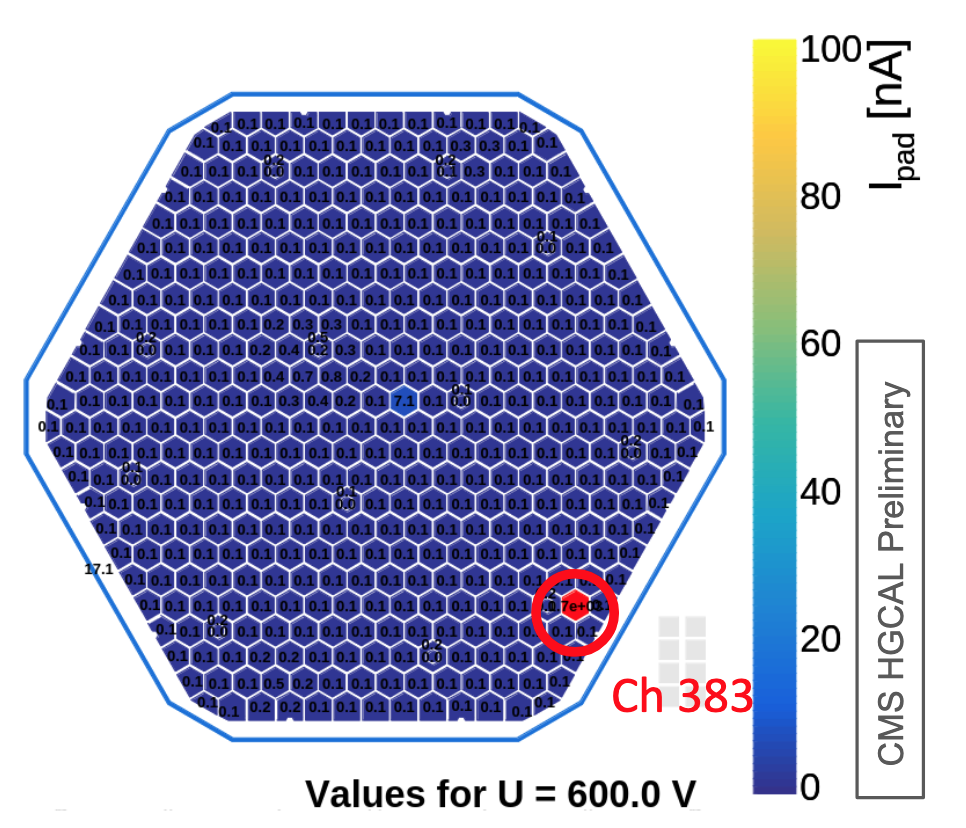}
    \caption{}
    \label{fig:stacking_cell}
  \end{subfigure}
  \hspace{0.001\textwidth} 
  \begin{subfigure}[b]{0.43\textwidth}
    \centering
    \includegraphics[width=\textwidth]{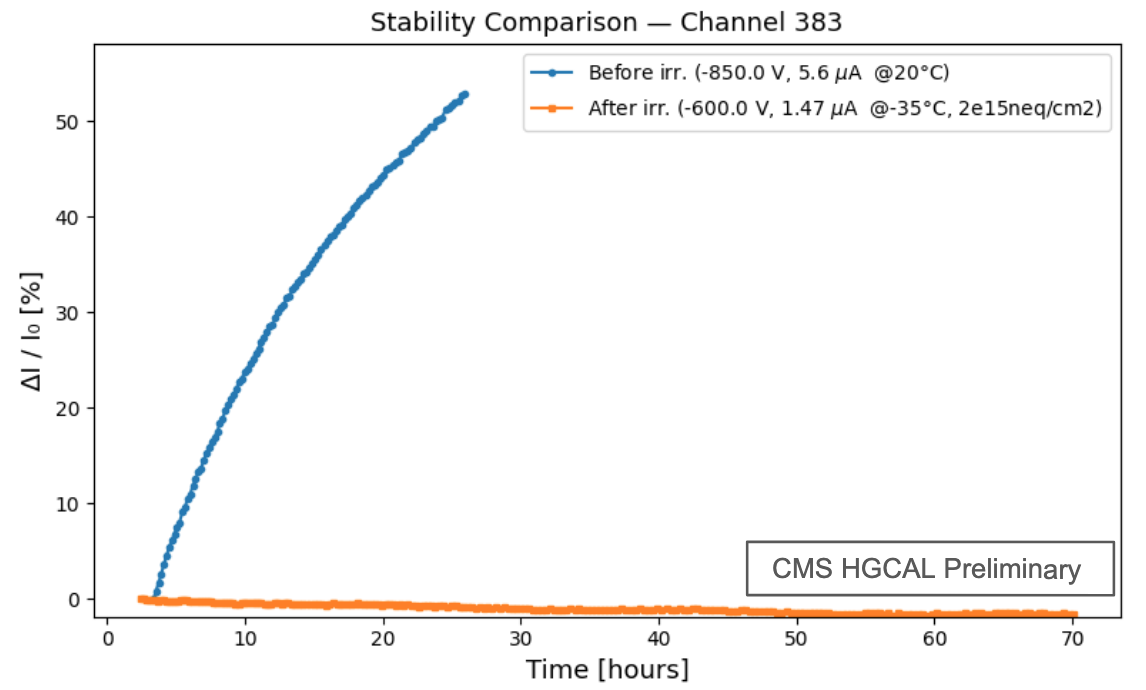}
    \caption{}
    \label{fig:current_ratio}
  \end{subfigure}
\vspace{-6pt}
  \caption{(a) The stacking-fault cell (383) before irradiation shows very high
    leakage current values ($1\,\mathrm{mA}$). (b) Ratio of current difference for each measurement compared to the first measurement versus time: before irradiation the ratio increases by 50\% in one day, while after irradiation
    the current decreases with time.}
  \label{fig:stacking}
\end{figure}

Epitaxial sensors containing stacking faults~\cite{airaksinen2005} were detected in approximately 5\% of
the sensors produced.
In cells with such defects, a high current --- accounting for a significant fraction
of the total current --- was observed (Figure~\ref{fig:stacking_cell}).
Before irradiation, stability runs revealed that the cell current increases with time
(Figure~\ref{fig:current_ratio}).
After irradiation, the leakage current is dominated by bulk radiation damage, and the
stacking-fault cell (383) reached the same values as the other cells.
Furthermore, after irradiation, the current of this cell decreased with time (Figure~\ref{fig:current_ratio}).
\vspace{-10pt}

\vspace{-14pt}
\section{Hot Regions after Irradiation}
\label{sec:hotregions}
\begin{figure}[htbp]
  \centering

  \begin{subfigure}[b]{0.35\textwidth}
    \centering
    \includegraphics[width=\textwidth]{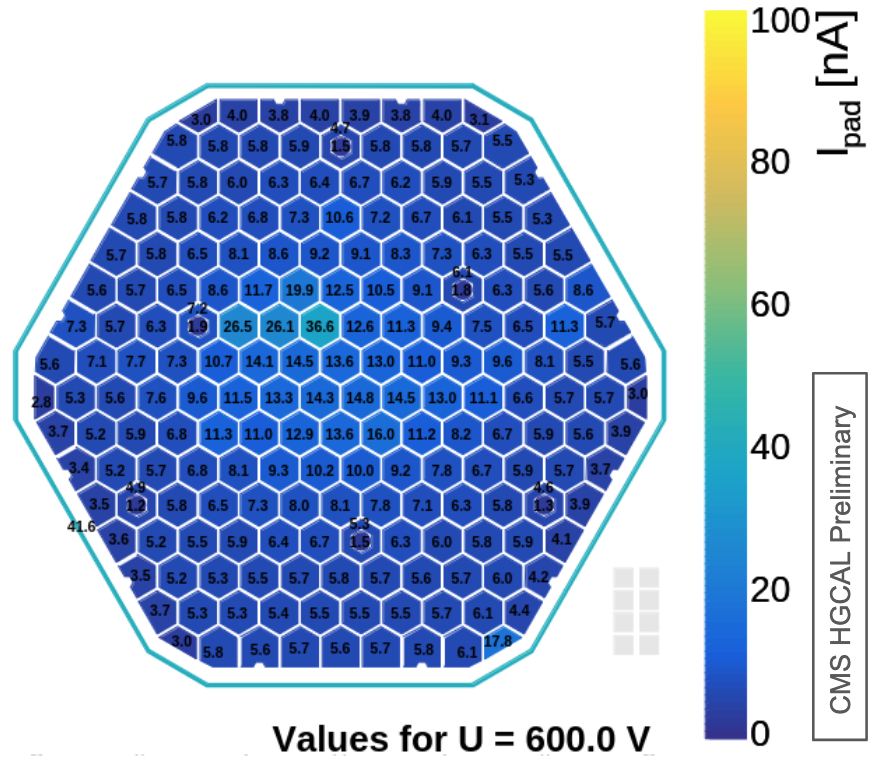}
    \caption{}
    \label{fig:hot_before}
  \end{subfigure}
  \hspace{0.001\textwidth} 
  \begin{subfigure}[b]{0.37\textwidth}
    \centering
    \includegraphics[width=\textwidth]{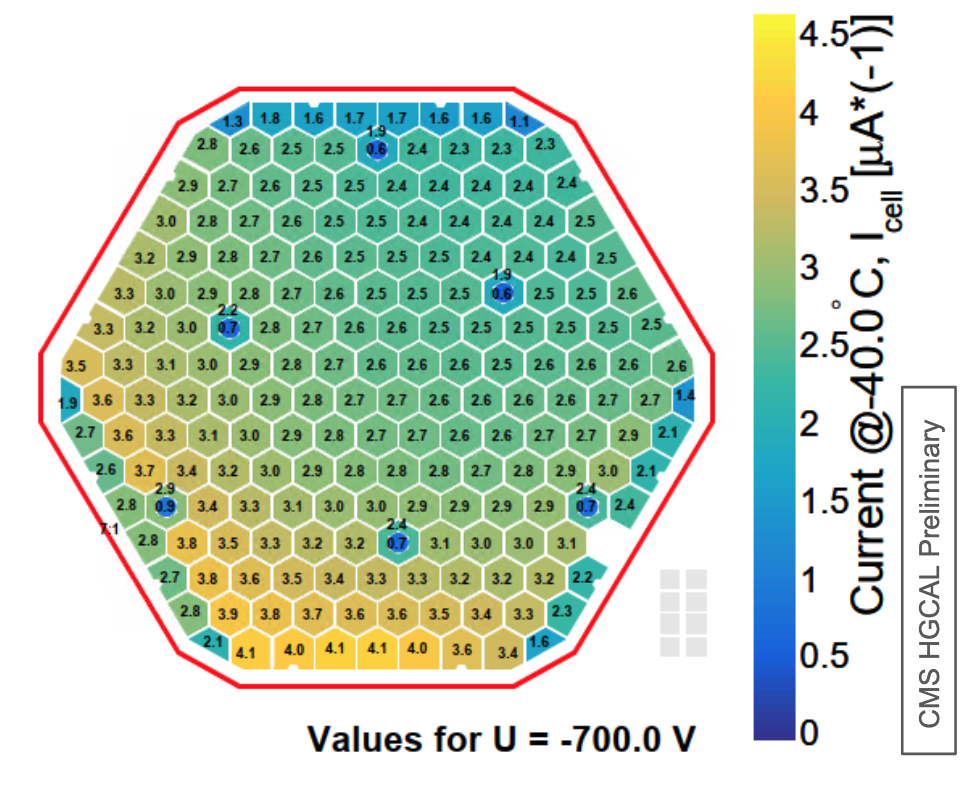}
    \caption{}
    \label{fig:hot_after}
  \end{subfigure}
\vspace{-6pt}
  \caption{(a) Elevated cell current for some channels before irradiation. (b) After
    irradiation, the leakage current is dominated by radiation damage and previously hot
    regions are no longer visible.}
  \label{fig:hot}
\end{figure}

\vspace{-9pt}
Many unirradiated HGCAL silicon sensors exhibit regions of elevated (Figure~\ref{fig:hot_before}) but moderate cell current
($<100$\,nA at 600\,V)~\cite{4th-DRD3}. The origin of this phenomenon is under evaluation.
We investigated whether this problem persists after irradiation, and found that for
irradiated cells the leakage current is dominated by bulk radiation damage, making the
previously elevated regions no longer visible (Figure~\ref{fig:hot_after}).
\vspace{-11pt}

\section{Summary}
\label{sec:summary}

The production of silicon sensors for the CMS HGCAL has been carried out by HPK between 2023 and 2025, resulting in the fabrication of approximately 25000 sensors. A comprehensive QC program accompanied the production process. All sensors underwent full testing at the vendor, while an additional sample of about 8\% of the main sensors was tested independently by the CMS SQC centres to ensure consistency and reliability. The sensors that pass the SQC meet the HGCAL specifications and will be used in the detector. All sensors will undergo further verification during the module assembly stage, and the current production yield is approximately 97\%. To mitigate the effects of potential yield losses during assembly, spare sensors and additional module components are being procured.

Irradiation studies have been performed to evaluate the radiation tolerance of the sensors under conditions relevant to the HL-LHC environment. These studies characterise the fluence delivered by the RINSC irradiation facility and the annealing behaviour of HGCAL 8-inch p-type silicon epitaxial sensors, and compare the results with single-diode measurements. By exploiting the fluence profile, a large number of fluence points were obtained, enabling a much more detailed study of the fluence dependence than is possible with diode measurements alone, which so far include only three fluence values per thickness.

Preliminary analyses of the depletion voltage evolution using the Hamburg model show results that are consistent in both magnitude and trend with measurements obtained from single-diode irradiation studies. The irradiation campaign has also provided valuable lessons for future measurements. In particular, it was found that finer voltage steps in CV measurements will improve the precision of depletion voltage extraction and the stability of model fits. Prior to irradiation, some channels exhibited localized regions with elevated leakage current (``\textit{hot regions}''). After irradiation, however, the leakage current becomes dominated by radiation-induced damage, and previously-identified stacking faults and hot regions are no longer distinguishable.

\section*{Acknowledgments}
\label{sec:Acknowledgments}

The author would like to thank colleagues of the CMS Collaboration at CERN (EP-CMS-EC, EP-DT) for their support. Special thanks go to all those who contributed to the SQC measurement and analysis work and provided materials, including Dorukhan Boncukcu, Acelya Deniz Gungordu, and Ufuk Guney Tok.
The author also extends sincere thanks to Leena Diehl and Eva Sicking for providing information and significant support during irradiation and related activities, as well as to Nicholas Hinton and other colleagues at Brown University for their invaluable help during the irradiation campaigns.


\end{document}